\documentclass{revtex4}
\usepackage{graphicx}
\usepackage{dcolumn}
\usepackage{bm}     

\usepackage[T1]{fontenc}        
\usepackage[latin1]{inputenc}   
\usepackage{times}              
\usepackage{amssymb}            %
\usepackage{amsmath}            
\usepackage{bbm}                
\usepackage{stmaryrd}           
\usepackage{xspace}             
\usepackage{subfigure}

\preprint{}

\begin{document}
\title{Penning trap mass measurements on $^{99-109}$Cd with ISOLTRAP and
implications on the rp process}

\author{M.~Breitenfeldt}
 \altaffiliation[Corresponding address: ]
 {CERN, Physics Department, 1211 Geneva 23, Switzerland\\
 Electronic address: Martin.Breitenfeldt@cern.ch\\
 This publication comprises part of the PhD thesis of M. Breitenfeldt}
\affiliation{Institut f{\"u}r Physik, Ernst-Moritz-Arndt-Universit{\"a}t, 17487 Greifswald, Germany}

\author{G.~Audi}
\affiliation{CSNSM-IN2P3-CNRS, 91405 Orsay-Campus, France}

\author{D.~Beck}
\affiliation{GSI, Planckstra{\ss}e 1, 64291 Darmstadt, Germany}

\author{K.~Blaum}
\affiliation{Max-Planck-Institut f\"{u}r Kernphysik, 69117 Heidelberg, Germany}
\affiliation{Ruprecht-Karls-Universit\"at, Institut f\"ur Physik, 69120 Heidelberg, Germany}

\author{S.~George}
\affiliation{Max-Planck-Institut f\"{u}r Kernphysik, 69117 Heidelberg, Germany}
\affiliation{Institut f{\"u}r Physik, Johannes Gutenberg-Universit{\"a}t, 55128 Mainz, Germany}

\author{F.~Herfurth}
\affiliation{GSI, Planckstra{\ss}e 1, 64291 Darmstadt, Germany}

\author{A.~Herlert}
\affiliation{CERN, Physics Department, 1211 Geneva 23, Switzerland}

\author{A.~Kellerbauer}
\affiliation{Commiss European Communities, Joint Res Ctr, European Inst Transuranium Elements, 76125 Karlsruhe, Germany}

\author{H.-J.~Kluge}
\affiliation{GSI, Planckstra{\ss}e 1, 64291 Darmstadt, Germany}
\affiliation{Ruprecht-Karls-Universit\"at, Institut f\"ur Physik, 69120 Heidelberg, Germany}

\author{M.~Kowalska}
\affiliation{CERN, Physics Department, 1211 Geneva 23, Switzerland}

\author{D.~Lunney}
\affiliation{CSNSM-IN2P3-CNRS, 91405 Orsay-Campus, France}

\author{S.~Naimi}
\affiliation{CSNSM-IN2P3-CNRS, 91405 Orsay-Campus, France}

\author{D.~Neidherr}
\affiliation{Institut f{\"u}r Physik, Johannes Gutenberg-Universit{\"a}t, 55128 Mainz, Germany}

\author{H.~Schatz}
\affiliation{NSCl, Department of Physics and Astronomy, Michigan State University, MI 48824, East Lansing, USA}
\affiliation{Joint Institute for Nuclear Astrophysics, Michigan State University, MI 48824, East Lansing, USA}

\author{S.~Schwarz}
\affiliation{NSCl, Department of Physics and Astronomy, Michigan State University, MI 48824, East Lansing, USA}

\author{L.~Schweikhard}
\affiliation{Institut f{\"u}r Physik, Ernst-Moritz-Arndt-Universit{\"a}t, 17487 Greifswald, Germany}

\date{\today}
\begin{abstract}
Penning trap mass measurements on neutron-deficient Cd isotopes $^{99-109}$Cd have been performed with the ISOLTRAP mass spectrometer at ISOLDE/CERN, all with relative mass uncertainties below $3\cdot10^{-8}$. A new mass evaluation has been performed. The mass of $^{99}$Cd has been determined for the first time which extends the region of accurately known mass values towards the doubly magic nucleus $^{100}$Sn. The implication of the results on the reaction path of the rp process in stellar X-ray bursts is discussed. In particular, the uncertainty of the abundance and the overproduction created by the rp-process for the mass $A=99$ is demonstrated by reducing the uncertainty of the proton-separation energy of $^{100}$In $S_p(^{100}$In) by a factor of 2.5.
\end{abstract}

\pacs{
      21.10.Dr {Binding energies and masses},
      32.10.Bi {Atomic masses, mass spectra, abundances, and isotopes},
      26.30.Ca {Explosive burning in accreting binary systems (novae, x-ray bursts)},
      27.60.+j {90 = A = 149}
   } 

\maketitle

%
\section{Introduction}
\label{sec:introduction}

Penning ion traps are versatile tools used in many areas in atomic and nuclear physics. One application is high precision mass spectrometry of atomic nuclei which leads to important input data for, e.g., nuclear structure studies \cite{Blaum2006,Schweikhard2006}. Numerous results with very high precision have been reported from a number  of facilities around the world for short-lived radioactive nuclides (ISOLTRAP \cite{Mukherjee2008}, CPT \cite{Clark2003}, JYFLTRAP \cite{Jokinen2006},  LEBIT \cite{Schwarz2003}, SHIPTRAP \cite{Block2005}, and TITAN \cite{Dilling2006}) covering the whole chart of nuclides. This allows one to test mass models and to improve mass predictions of exotic nuclides which have not been addressed so far. In nuclear astrophysics mass differences and thus nuclear masses are essential for the modeling of many nucleosynthesis sites. A current goal is the extension of high-precision mass measurements to nuclei very far from stability, in particular towards the very neutron-deficient nuclei in the rapid proton capture process (rp process) and towards the very neutron-rich nuclei in the the rapid neutron capture process (r process). This goal is also addressed by storage ring mass spectrometry at the ESR facility at GSI \cite{Franzke2008}. ISOLTRAP has recently contributed a number of precision mass measurements to this area such as $^{22}$Mg \cite{Mukherjee2004,Mukherjee2008b} and $^{72}$Kr \cite{Rodriguez2004,Rodriguez2005} on the neutron-deficient side, and $^{80,81}$Zn \cite{Baruah2008}, $^{95}$Kr \cite{Delahaye2006} and $^{132,134}$Sn \cite{Sikler2005,Dworschak2008} on the neutron-rich side.

In this paper we present Penning trap mass measurements of neutron-deficient Cd isotopes out to $^{99}$Cd that are important for modeling
the isotopic abundances produced by the astrophysical rp process \cite{WaW81,Schatz98,Frohlich:2005ys,Pruet:2005qd}. The rp process is a sequence of rapid proton captures and $\beta^+$ decays, often close to the proton drip line. For the $A\approx 99$ mass region, the rp process has been suggested \cite{WaW81,Schatz98} and discussed \cite{Fallis2008} as a candidate to explain the long-standing puzzle of the origin of the relatively large amounts of $^{92,94}$Mo and $^{96,98}$Ru in the solar system \cite{Lodders2003}. These form a lower-abundance group of so-called "p nuclei" that are shielded from neutron capture in the s and r processes, which synthesize the rest of the heavy elements in nature. While standard p-process scenarios based on photodisintegration processes produce most other p-nuclei, they severely underproduce  $^{92,94}$Mo and $^{96,98}$Ru \cite{Rapp:2006rd,Arn03}.

The rp process is the main energy source of type I X-ray bursts on the surface of accreting neutron stars \cite{Schatz2006}.
In some bursts characterized by long timescales of the order of 100\,s the rp process
can reach the Cd region \cite{Schatz2001}. A reliable estimate of the
produced composition is needed to model neutron star crust processes
that are related to a number of observables such as the rare superbursts
or the cooling of transiently accreting neutron stars \cite{Gupta2007}. In addition,
it has been shown that a small fraction of the processed matter could be
ejected during X-ray bursts, renewing interest in these scenarios in terms of
producing the Mo and Ru p-isotopes \cite{Weinberg2005}.

The rp process is also thought to occur in proton-rich neutrino-driven outflows
in core collapse supernovae \cite{Frohlich:2005ys,Pruet:2005qd}. Because of the prominent role that neutrinos
play in this nucleosynthesis it is referred to as "$\nu$p process". It has been
shown that for certain model parameters the process can synthesize the
Mo and Ru p-isotopes and that it passes through the $^{99}$Cd
region investigated in this work \cite{Pruet:2005qd}. For both scenarios the importance of accurate nuclear masses has been
discussed before \cite{Pruet:2005qd,Fallis2008,Schatz_Rehm2006,Fisker:2007ay,Weber:2008pb,Weber2008_1}.

\section{Setup and Procedure}
\label{sec:setup}
\begin{figure}[t]
    \begin{center}
        \includegraphics[width=80mm]{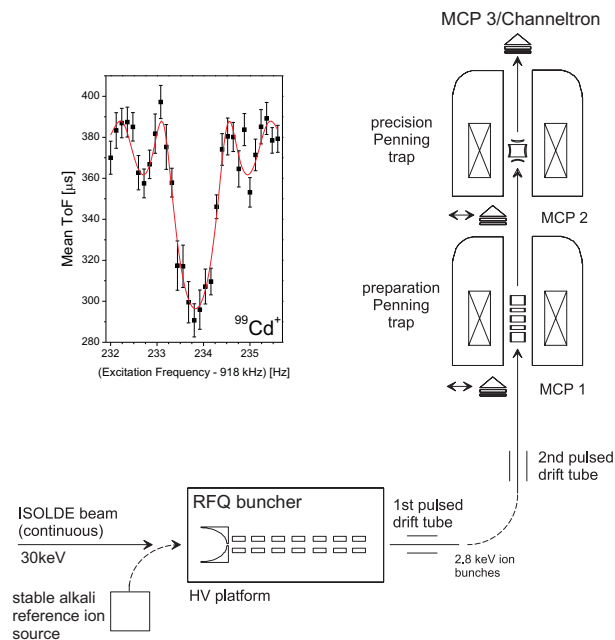}
    \end{center}
    \caption{\label{fig:setup}The triple-trap mass spectrometer ISOLTRAP with the three main parts: a RFQ buncher and two Penning traps. The inset shows a typical time-of-flight ion-cyclotron resonance for $^{99}$Cd$^+$ with a fit of the theoretical line-shape (solid line) to the data \protect\cite{Koenig1995}.}
\end{figure}

The measurements have been performed at the triple-trap mass spectrometer ISOLTRAP \cite{Mukherjee2008} at the isotope separator ISOLDE \cite{Kugler2000} at CERN, Geneva. As shown in Fig.~\ref{fig:setup} ISOLTRAP consists of three main parts: a linear radio frequency quadrupole (RFQ) buncher \cite{Herfurth2001,Herfurth2003} for accumulation of the ions, a gas-filled cylindrical Penning trap for cooling, centering and mass separation of the ions \cite{Savard1991} and a hyperbolical Penning trap in ultra-high vacuum for the determination of the cyclotron frequency $\nu_c$. The present status of the experimental setup is described in more detail in \cite{Mukherjee2008}.

In this work the Cd isotopes were created by 1.4-GeV proton pulses impinging on a Sn liquid-metal target with a thickness of 115\,g\,cm$^{-2}$. After evaporation from the target the cadmium atoms were ionized in a FEBIAD hot plasma ion source \cite{Sundell1992}, accelerated to 30\,kV, sent through the General Purpose Separator (GPS) with a resolving power of $m/\Delta m$~=~800, and transported to the ISOLTRAP experiment.

\begin{figure}[h]
    \begin{center}
        \includegraphics[width=80mm]{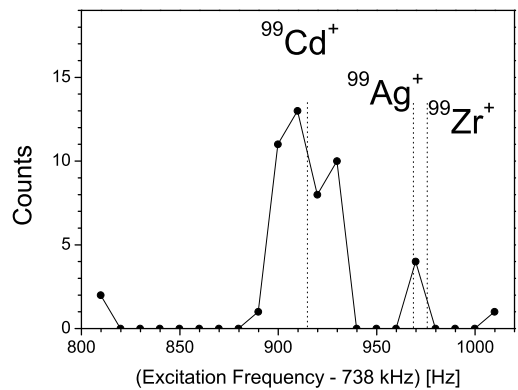}
    \end{center}
    \caption{\label{fig:cooling_resonance} A cooling resonance for $^{99}$Cd$^+$ in the preparation trap. The number of ions observed after ejection is plotted as a function of the excitation frequency $\nu_{rf}$. $^{99}$Cd$^+$ is centered at about 738.92\,kHz. Dashed lines indicate the positions of the cyclotron frequencies of $^{99}$Ag$^+$ and $^{99}$Zr$^+$, respectively.}
\end{figure}

At ISOLTRAP the ions were accumulated and cooled in the RFQ~buncher \cite{Herfurth2003}, which was elevated to a potential of 30\,kV to decelerate the incoming continuous radioactive ion beam. The ions were ejected with a bunch length of about 1\,$\mu$s and sent to the preparation Penning trap where the buffer-gas cooling technique \cite{Savard1991} with a resolving power of about 20000 was applied for isobaric purification. Figure~\ref{fig:cooling_resonance} shows an example of a cooling resonance for $^{99}$Cd$^+$: The number of detected ions after centering is plotted as a function of the quadrupolar rf excitation frequency. The central peak corresponds to $^{99}$Cd$^+$, while the small peak to the higher-frequency side corresponds to the frequency of $^{99}$Ag$^+$. ($^{99}$Zr$^+$ would appear at almost the same cyclotron frequency as $^{99}$Ag$^+$, but is not expected to be released from the target.) Subsequently the ions were transferred to the precision Penning trap for the determination of the cyclotron frequency $\nu_c$ using the Time-of-Flight Ion-Cyclotron-Resonance (ToF-ICR) method \cite{Graeff1980,Bollen1990}. The value for $\nu_c$ was obtained by fitting the theoretical line shape of the ToF-ICR to the data \cite{Koenig1995}.

In the case of $^{103}$Cd a possible $^{103}$Mo contamination at $m/\Delta m =480000$ was excluded by the application of a corresponding dipolar excitation at the reduced cyclotron frequency of the contaminant in the precision trap, which leads to radial ejection \cite{Roosbroeck2004}. In all other cases the masses of possible contaminants are sufficiently far away from the masses of the nuclides of interest to eliminate them during cyclotron cooling in the preparation trap.

The measured cyclotron resonances were investigated with respect to possible shifts due to the presence of simultaneously stored isobaric ions by the standard analysis procedure applied at ISOLTRAP \cite{Mukherjee2008,Kellerbauer2003}. No indication for any contamination was found. This procedure has repeatedly demonstrated that uncertainties down to $2\cdot10^{-8}$ are possible and reproducible with ISOLTRAP \cite{Guenaut2007,Yazidjian2007}.


\begin{table}
    \center
    \caption{Half-lives and cyclotron frequency ratios $r=\nu_c(^{85}$Rb$^+)/\nu_c(^{\text{A}}$Cd$^+)$ between the reference nuclide $^{85}$Rb and the neutron-deficient cadmium nuclides $^{99-109}$Cd.}
    \label{tab:frequency ratio}
        \begin{tabular}{ccc}
            \hline\noalign{\smallskip}
            Nuclide & Half-life & $r=\nu_c\left(^{85}\text{Rb}^+\right)/\nu_c\left(^{\text{A}}\text{Cd}^+\right)$\\
            \noalign{\smallskip}\hline\noalign{\smallskip}
            $^{99}$Cd  & 16(3)\,s     &  1.165\,032\,756\,0(202)\\
            $^{100}$Cd & 49.1(0.5)\,s &  1.176\,755\,855\,2(208)\\
            $^{101}$Cd & 1.36(5)\,min &  1.188\,512\,101\,2(189)\\
            $^{102}$Cd & 5.5(0.5)\,min&  1.200\,240\,767\,7(218)\\
            $^{103}$Cd & 7.3(0.1)\,min&  1.212\,005\,235\,3(250)\\
            $^{104}$Cd &57.7(1.0)\,min&  1.223\,740\,297\,6(228)\\
            $^{105}$Cd &55.5(0.4)\,min&  1.235\,512\,679\,7(182)\\
            $^{106}$Cd & stable       &  1.247\,254\,328\,2(215)\\
            $^{107}$Cd & 6.50(2)\,h   &  1.259\,033\,102\,3(225)\\
            $^{108}$Cd & stable       &  1.270\,781\,503\,2(270)\\
            $^{109}$Cd & 461.4(1.2)\,d&  1.282\,567\,977\,2(219)\\
            \noalign{\smallskip}\hline
        \end{tabular}
\end{table}

\section{Experimental results}
\label{sec:results}

Over a period of five days between three and five resonances for each of the eleven investigated nuclides $^{99-109}$Cd have been recorded. The inset of Fig.~\ref{fig:setup} shows a typical example for a ToF-ICR curve of $^{99}$Cd$^+$. The magnetic field strength is interpolated between two reference measurements of $^{85}$Rb$^+$. The averaged values of cyclotron-frequency ratios $r$ between the reference nuclide $^{85}$Rb and the neutron-deficient cadmium isotopes $^{99-109}$Cd, $r=\nu_c(^{85}$Rb$^+)/\nu_c(^{\text{A}}$Cd$^+)$, are given in Table~\ref{tab:frequency ratio}.

\begin{table*}
    \center
    \caption{The mass excess ($ME$) of the neutron deficient Cd isotopes with $A=99-109$ for the measurements performed at ISOLTRAP (this work), SHIPTRAP \protect\cite{Martin2007}, and JYFLTRAP \protect\cite{Elomaa2008}. The adjusted JYFLTRAP $ME$ values calculated from the frequency ratios published in \protect\cite{Elomaa2008} using a reference from the current AME are given in the last column.}
    \label{tab:mass excess}
        \begin{tabular}{ccr@{$($}lcc}
            \hline\noalign{\smallskip}
            Nuclide & $ME$(ISOLTRAP)& \multicolumn{2}{c}{$ME$(SHIPTRAP)}& $ME$(JYFLTRAP publ.)& $ME$(JYFLTRAP adj.)\\
                    & / keV         & \multicolumn{2}{c}{/keV}          &   / keV             &    / keV           \\
            \noalign{\smallskip}\hline\noalign{\smallskip}
            $^{99}$Cd  & -69931.1(1.6) & \multicolumn{2}{c}{}           &                     &                    \\
            $^{100}$Cd & -74194.6(1.6) & \multicolumn{2}{c}{}           &                     &                    \\
            $^{101}$Cd & -75836.4(1.5) & -75849&10)                     & -75827.8(5.6)       & -75831.2(5.1)      \\
            $^{102}$Cd & -79659.6(1.7) & -79672&7)                      & -79655.6(5.3)       & -79659.1(4.8)      \\
            $^{103}$Cd & -80651.2(2.0) & -80651&10)                     & -80648.5(5.3)       & -80652.0(4.8)      \\
            $^{104}$Cd & -83968.5(1.8) & -83979&5)                      & -83962.9(5.6)       & -83966.4(5.0)      \\
            $^{105}$Cd & -84334.0(1.4) & \multicolumn{2}{c}{}           & -84330.1(5.5)       & -84333.8(4.8)      \\
            $^{106}$Cd & -87130.4(1.7) & \multicolumn{2}{c}{}           &                     &                    \\
            $^{107}$Cd & -86990.4(1.8) & \multicolumn{2}{c}{}           &                     &                    \\
            $^{108}$Cd & -89252.7(2.1) & \multicolumn{2}{c}{}           &                     &                    \\
            $^{109}$Cd & -88503.7(1.7) & \multicolumn{2}{c}{}           &                     &                    \\
            \noalign{\smallskip}\hline
        \end{tabular}
\end{table*}

\begin{figure}[ht]
  \begin{center}
      \subfigure{{\includegraphics[width=8cm]{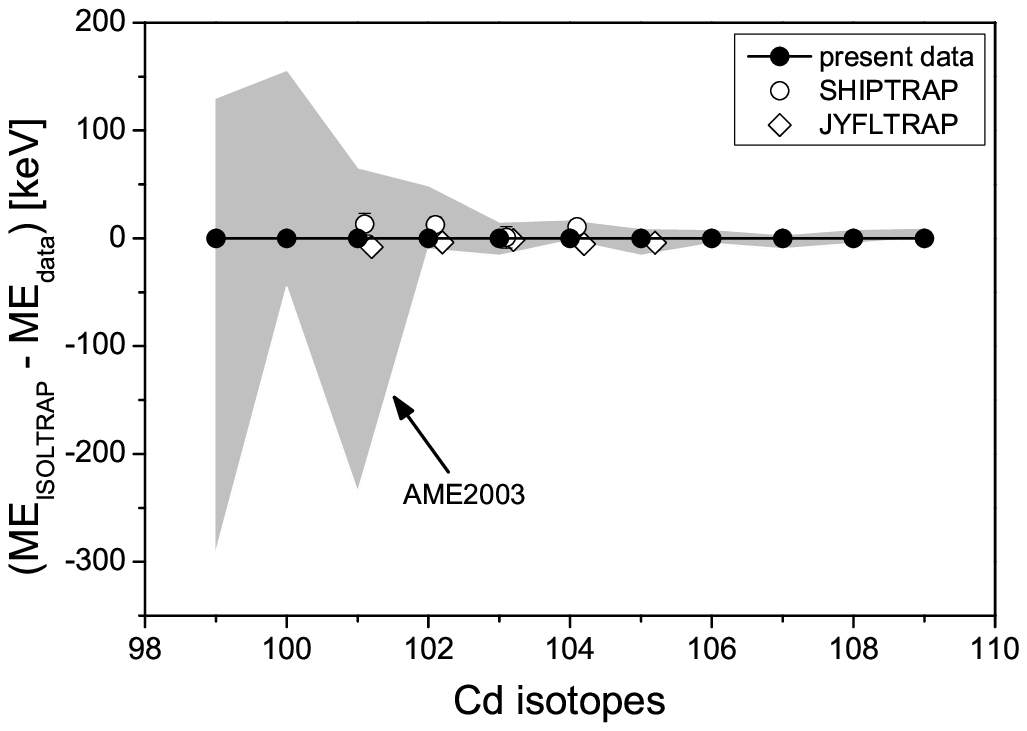}}\label{fig:ME_large}}\qquad
      \subfigure{{\includegraphics[width=8cm]{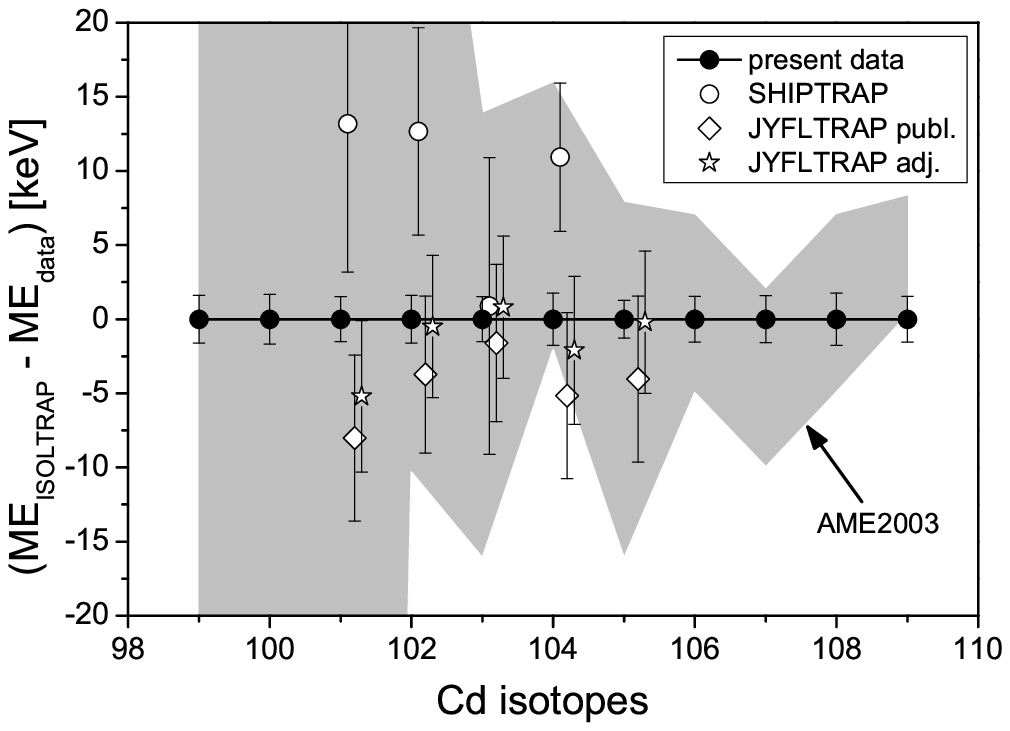}}\label{fig:ME small}}\qquad
    \caption{\label{fig:ME} Top: Differences between the new mass-excess values measured at ISOLTRAP (full circles) and those from AME2003 \protect\cite{Audi2003} and from SHIPTRAP \protect\cite{Martin2007} (open cicles) and JYFLTRAP \protect\cite{Elomaa2008} (open squares). The new ISOLTRAP masses were chosen as a reference. The shaded area represents AME2003 values. Bottom: Vertical zoom of top figure including recalculated values from JYFLTRAP using the mass of $^{96}$Mo from the most recent AME (stars).}
  \end{center}
\end{figure}

As shown in Fig.~\ref{fig:ME}, the measurements performed at ISOLTRAP (full symbols) agree with the literature values of the latest Atomic-Mass Evaluation (AME2003) \cite{Audi2003} within the uncertainties. Note that the mass of $^{99}$Cd was determined experimentally for the first time. This plot also contains the recent mass excess values obtained by SHIPTRAP at GSI \cite{Block2005} and by JYFLTRAP at IGISOL \cite{Jokinen2006}. In these campaigns the masses of $^{101-105}$Cd have been determined \cite{Martin2007,Elomaa2008}, as listed in Table~\ref{tab:mass excess} and plotted as open symbols in Fig.~\ref{fig:ME}. In the case of the SHIPTRAP measurements a tendency to higher mass excess values is observed. A new mass evaluation has been performed for this paper in order to present the full impact of these, and related results from the same region.  The new evaluation follows exactly the same procedure as that outlined in the AME2003 \cite{Wapstra2003}, but using the updated flow-of-information matrices. A first evaluation was calculated including the values of SHIPTRAP and JYFLTRAP and a second including also ISOLTRAP. New averaged values are obtained, which are given in the last two columns of Table~\ref{tab:mass excess adjusted} and demonstrate the influence of the ISOLTRAP data.

\section{Discussion}
\label{sec:discussion}

\begin{table*}
    \center
    \caption{The mass excess ($ME$) of the neutron deficient Cd isotopes with $A=99-109$ for the measurements performed at ISOLTRAP (this work), those listed in the AME2003 \protect\cite{Audi2003}, those obtained in an atomic-mass evaluation before the ISOLTRAP data entered (including SHIPTRAP \protect\cite{Martin2007} and JYFLTRAP data \protect\cite{Elomaa2008}) and the newly adjusted values (last column). The symbol \# marks the AME value of $^{99}$Cd as extrapolated from systematics.}
    \label{tab:mass excess adjusted}
        \begin{tabular}{ccr@{$($}lr@{$($}lc}
            \hline\noalign{\smallskip}
            Nuclide & $ME$(ISOLTRAP)& \multicolumn{2}{c}{$ME$(AME2003)}& \multicolumn{2}{c}{$ME$(AME before)}   & $ME$(AME after)\\
                    & / keV         & \multicolumn{2}{c}{/keV}&\multicolumn{2}{c}{/keV}&    / keV       \\
            \noalign{\smallskip}\hline\noalign{\smallskip}
            $^{99}$Cd  & -69931.1(1.6) &-69850&210)\#& -69850&210)\#    & -69931.1(1.6)  \\
            $^{100}$Cd & -74194.6(1.6) & -74250&100) & -74252&65)       & -74194.6(1.7)  \\
            $^{101}$Cd & -75836.4(1.5) & -75750&150) & -75835.8&4.8)    & -75836.0(1.4)  \\
            $^{102}$Cd & -79659.6(1.7) & -79678&29)  & -79664.4&4.1)    & -79659.5(1.7)  \\
            $^{103}$Cd & -80651.2(2.0) & -80649&15)  & -80656.3&4.2)    & -80652.0(1.8)  \\
            $^{104}$Cd & -83968.5(1.8) & -83975&9)   & -83968.7&4.7)    & -83968.3(1.6)  \\ 
            $^{105}$Cd & -84334.0(1.4) & -84330&12)  & -84334.4&4.9)    & -84333.8(1.3)  \\ 
            $^{106}$Cd & -87130.4(1.7) & -87132&6)   & -87128.2&5.0)    & -87130.4(1.7)  \\ 
            $^{107}$Cd & -86990.4(1.8) & -86985&6)   & -86986.3&5.7)    & -86990.1(1.7)  \\ 
            $^{108}$Cd & -89252.7(2.1) & -89252&6)   & -89251.9&5.5)    & -89252.6(2.1)  \\ 
            $^{109}$Cd & -88503.7(1.7) & -88508&4)   & -88508.2&3.4)    & -88504.7(1.6)  \\ 
            \noalign{\smallskip}\hline
        \end{tabular}
\end{table*}

\subsection{Mass Evaluation}
\label{sec:mass_evaluation}
\begin{figure*}[ht]
  \begin{center}
      \subfigure[]{\rotatebox{270}{\includegraphics[width=8cm]{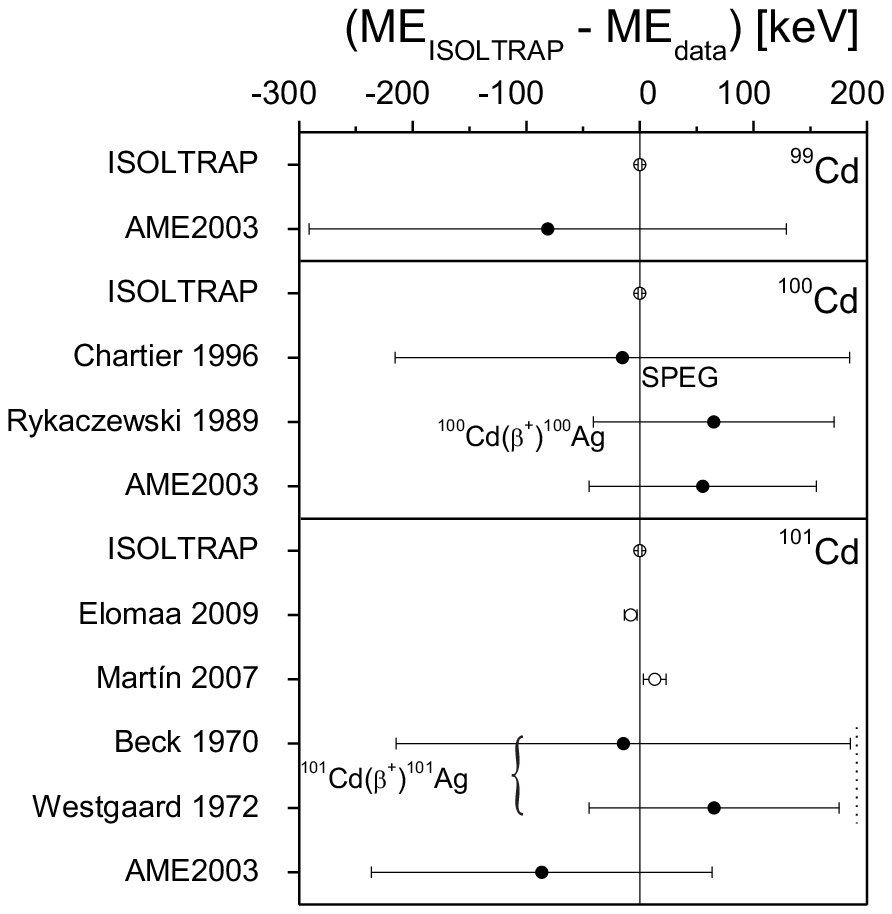}}\label{fig:AME2003:Cd99to101}}\qquad
      \subfigure[]{\rotatebox{270}{\includegraphics[width=8cm]{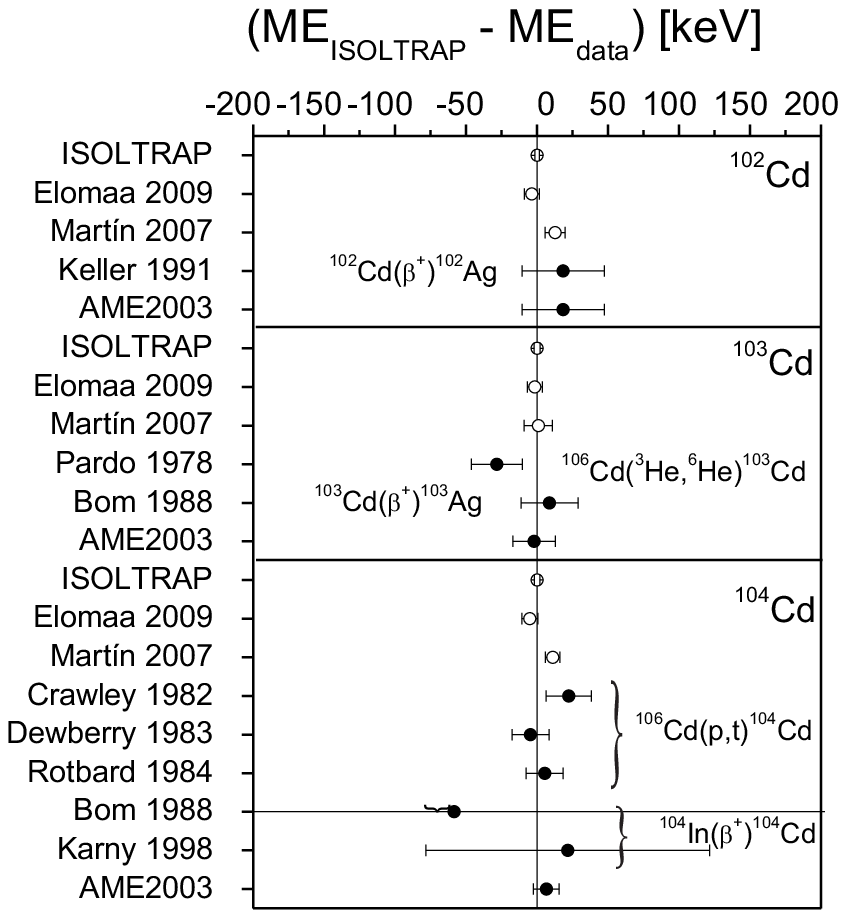}}\label{fig:AME2003:Cd102to104}}\qquad
      \subfigure[]{\rotatebox{270}{\includegraphics[width=8cm]{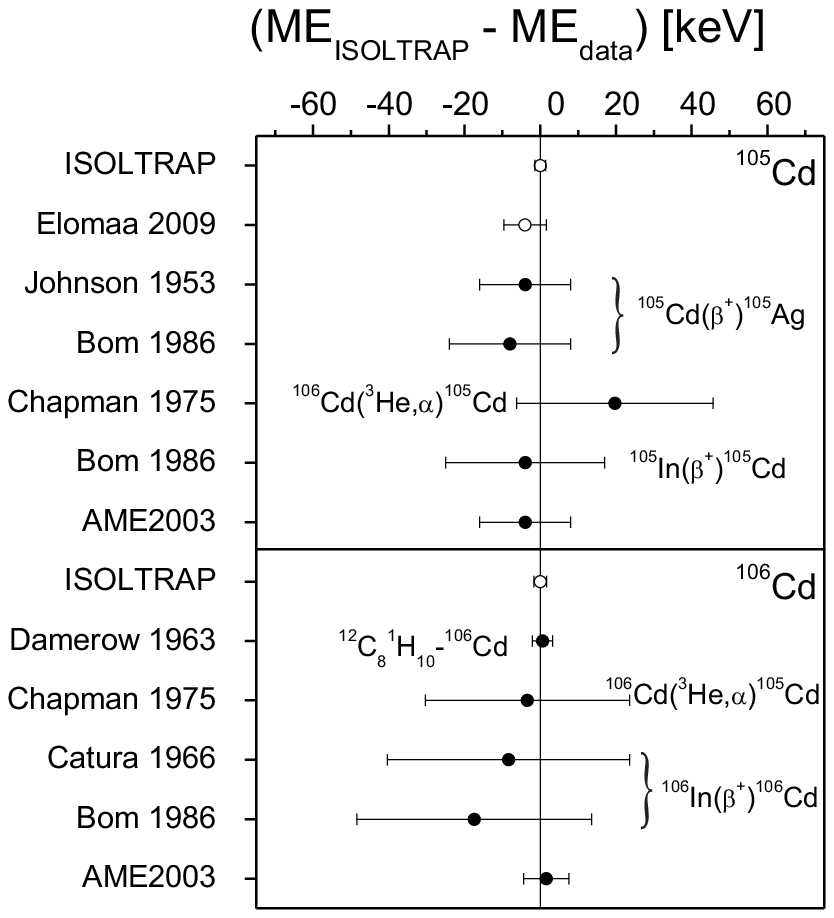}}\label{fig:AME2003:Cd105to106}}\qquad
      \subfigure[]{\rotatebox{270}{\includegraphics[width=8cm]{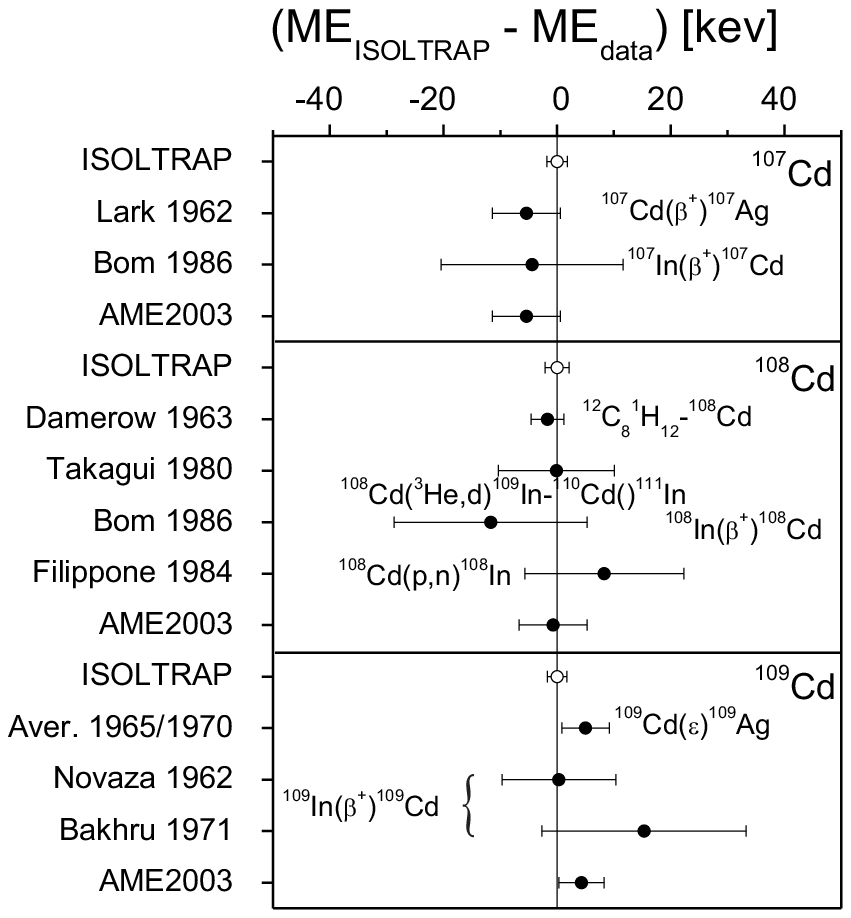}}\label{fig:AME2003:Cd107to109}}\qquad
    \caption{\label{fig:AME2003:Cd99-109} Comparison of the mass-excess values of ISOLTRAP with the data of the Penning trap setups (SHIPTRAP \protect\cite{Martin2007}, JYFLTRAP \protect\cite{Elomaa2008}), the data which have been included in the AME2003 \protect\cite{Audi2003}, and with the resulting AME2003 values for the nuclides $^{99-109}$Cd. The braces connect similar reactions/experiments. The Penning trap values are marked with open circles, the older experimental data and the AME2003 values are indicated with full circles.}
  \end{center}
\end{figure*}

In the following the results obtained in this work are compared to previous data which were available for the Atomic-Mass Evaluation in 2003 \cite{Audi2003}. In Fig.\,\ref{fig:AME2003:Cd99-109} differences between mass-excess values obtained from ISOLTRAP and from the other two Penning trap experiments and the AME2003 are plotted as well as from mass-excess values calculated from the input data of the AME2003.

The SHIPTRAP data were already included in the mass evaluation, as published by Mart\'{\i}n \textit{et al.}  \cite{Martin2007}. The JYFLTRAP frequency ratios from \cite{Elomaa2008} were included in the present evaluations as given in Tab.\,\ref{tab:mass excess adjusted} in the last two columns. Using the ISOLTRAP frequency ratios a new atomic-mass evaluation was performed to check the influence of the new data on the AME network (Tab.\,\ref{tab:mass excess adjusted} last column). The individual cases are discussed in the following sections but there is an important general observation: Since the last published evaluation (in 2003 \cite{Audi2003}), the masses of many nuclides have changed. One of these is $^{96}$Mo, the reference mass used by JYFLTRAP to derive the masses in \cite{Elomaa2008}, which moved by 3.2\,keV. Elomaa \textit{et al.} \cite{Elomaa2008} reported deviations of 1.8$\sigma$ - 2.1$\sigma$ from the SHIPTRAP mass values ($^{101,102,104}$Cd). When these masses are recalculated using the new $^{96}$Mo mass value, the new JYFLTRAP values are in perfect agreement with those of ISOLTRAP and the deviation from the SHIPTRAP values is reduced to slightly over 1$\sigma$ (see Fig.\,\ref{fig:ME small} and Tab.\,\ref{tab:mass excess}).

\begin{table}
    \center
    \caption{The influences of the experimental data from ISOLTRAP (this work) and from JYFLTRAP \protect\cite{Elomaa2008} on the current AME on the mass excess values of $^{A}$Cd and $^{96}$Mo. The given influences of the SHIPTRAP data \protect\cite{Martin2007} are hypothetical, these data have not been included due to their low significance.}
    \label{tab:influences}
        \begin{tabular}{ccccc}
            \hline\noalign{\smallskip}
            Nuclide & \multicolumn{4}{c}{Influences of experimental data}\\
                    & \multicolumn{3}{c}{on the Cd nuclides}        & on $^{96}$Mo\\
                    & ISOLTRAP  & SHIPTRAP\footnote{hypothetically} & JYFLTRAP & JYFLTRAP\\
            \noalign{\smallskip}\hline\noalign{\smallskip}
            $^{99}$Cd  & 100\%  &                                   &          &         \\
            $^{100}$Cd & 100\%  &                                   &          &         \\
            $^{101}$Cd & 92.9\% &           2\%\footnotemark[1]     &   7.1\%  &    9.0\%\\
            $^{102}$Cd & 89.4\% &           6\%\footnotemark[1]     &   10.6\% &    9.9\%\\
            $^{103}$Cd & 84.7\% &           3\%\footnotemark[1]     &   12.3\% &    9.9\%\\
            $^{104}$Cd & 90.3\% &          10\%\footnotemark[1]     &   9.7\%  &    9.3\%\\
            $^{105}$Cd & 92.9\% &                                   &   6.4\%  &   10.2\%\\
            $^{106}$Cd & 99.7\% &                                   &          &         \\
            $^{107}$Cd & 91.5\% &                                   &          &         \\
            $^{108}$Cd & 94.0\% &                                   &          &         \\
            $^{109}$Cd & 82.9\% &                                   &          &         \\
            \noalign{\smallskip}\hline
        \end{tabular}
\end{table}

The reasons for the $^{96}$Mo mass change are multiple, mostly related to the removal or replacement of conflicting data that were linked to $^{96}$Mo (causing a -3.2-keV shift between the AME2003 and the new AME). The question of links is a key point here. It is important to remember that it is not a mass that is measured in a trap, but a cyclotron frequency ratio i.e., a link between two nuclides. As the reference a nuclide is chosen that already has a small uncertainty in its mass. In the case of $^{96}$Mo, the uncertainty was 1.9\,keV. As JYFLTRAP reported several frequency ratios involving this nuclide, the ensemble of these links also contributed to a reduction in the $^{96}$Mo uncertainty (to 1.5\,keV) as well as the remaining 0.1\,keV shift. Hence this is a case which illustrates the importance of the mass evaluation. For this reason the following discussion refers to JYFLTRAP data recalculated with the new $^{96}$Mo mass instead of to the published values \cite{Elomaa2008}, in order to avoid conflicts, which are already solved in the present adjustment.

Like JYFLTRAP, the SHIPTRAP measurements contribute only slightly to the final mass results as compared to ISOLTRAP. In the AME, there is a distinction between "influence" (how much a datum affects a particular mass) and "significance" (how much a datum affects all the table). It is the policy of the AME that only data having a "significance" of more than one ninth are used in the flow-of-information matrix \cite{Wapstra2003}. This minimizes the propagation of inaccurate data with no sacrifice in overall precision.

The SHIPTRAP data \cite{Martin2007}, obtained by measuring the link $^{85}$Rb - $^A$Cd, contribute less than the cut-off criterion for the case of the cadmium mass values as given in Tab.\,\ref{tab:influences}. Moreover, they have no influence on the value of $^{85}$Rb as it was measured by \cite{Bradley1999} to an accuracy of about 11\,eV. Thus, the "significance" of the SHIPTRAP data is concentrated on the mass being investigated. As a consequence, the SHIPTRAP data shown in Tab.\,\ref{tab:mass excess} are excluded from the evaluation.

\begin{figure}[ht]
  \begin{center}
      {{\includegraphics[width=8cm]{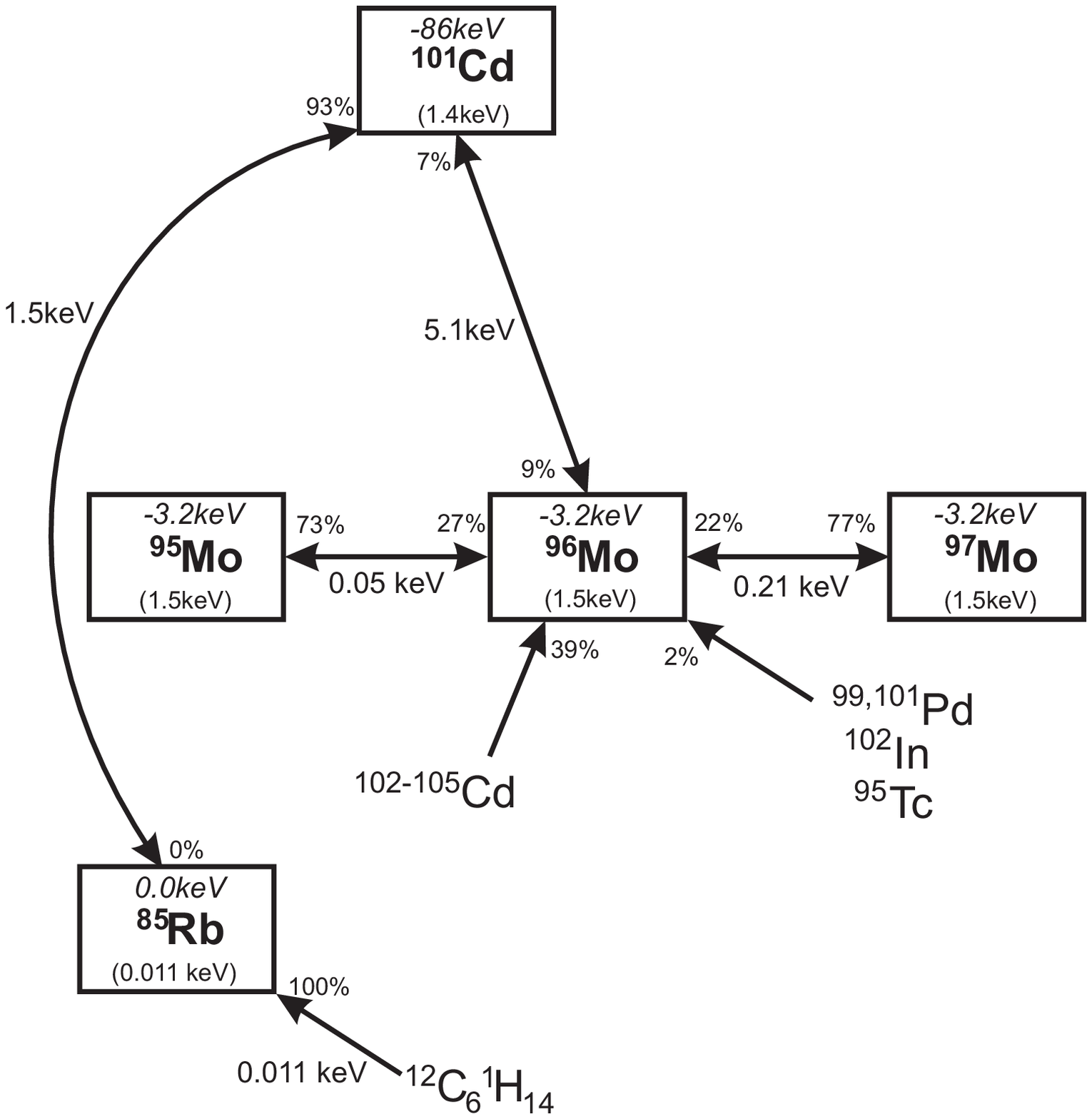}}}\qquad
    \caption{\label{fig:evaluations} A zoom into the network of experimental results and mass values on the nuclides $^{101}$Cd and $^{96}$Mo for the current evaluation. The boxes represent nuclides with their uncertainties. The \textit{italic} numbers indicated the shift of the mass excess value as compared to the AME2003. The connections between the boxes with the arrow-heads show the influences of the data on the certain masses. The values close to the links represent the uncertainties of the data. In the case of frequency ratios the experimental results are linearized, to do matrix calculations \protect\cite{Audi2003}. The data with low influence on $^{96}$Mo connecting to $^{99,101}$Pd, $^{102}$In, and $^{95}$Tc are mainly  determining the other end of the link and thus above the limit for insignificance.}
  \end{center}
\end{figure}

This is different for the data from JYFLTRAP. As can be seen from Tab.\,\ref{tab:influences}, the JYFLTRAP data have low influence on the cadmium mass values. However, JYFLTRAP has investigated the link $^{96}$Mo - $^{A}$Cd. As the mass value of $^{96}$Mo was previously only known to 1.9\,keV. Thus, there is also a flow of information from the JYFLTRAP data towards $^{96}$Mo. The "influence" of the JYFLTRAP data reduces the uncertainty of the $^{96}$Mo mass value to 1.5\,keV as shown in Fig.\,\ref{fig:evaluations} and therefor the "significance" of the JYFLTRAP data is increased. Therefor these data are included in the evaluation.

\begin{figure}[ht]
  \begin{center}
      \subfigure[]{\rotatebox{270}{\includegraphics[width=8cm]{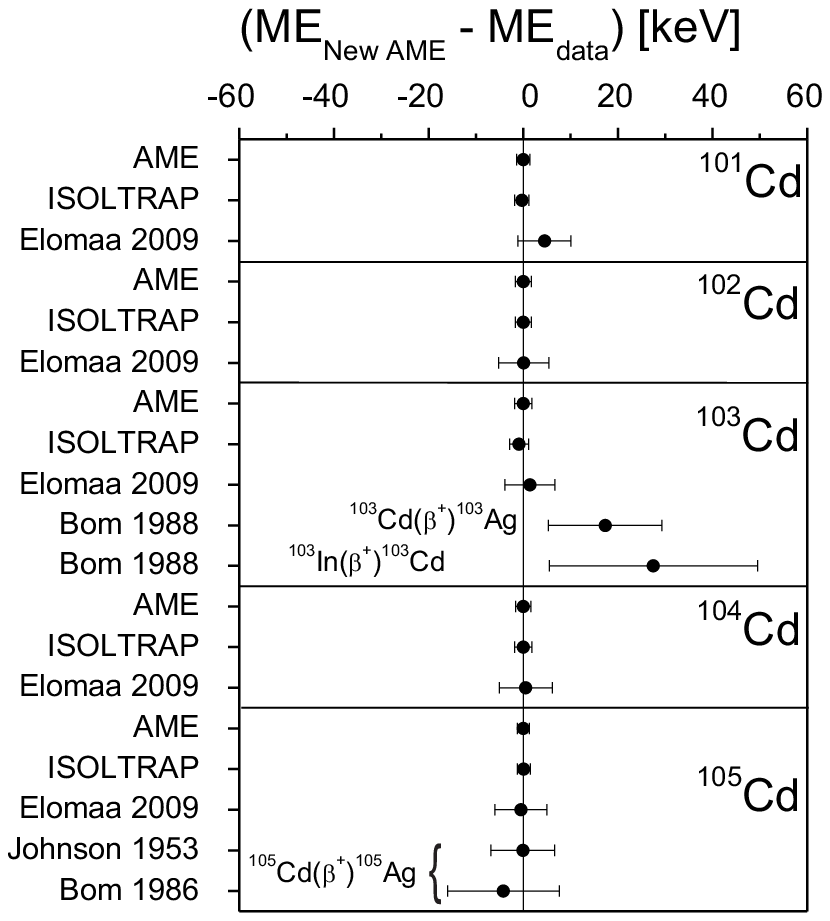}}\label{fig:NEW_AME:Cd103to106}}\qquad
      \subfigure[]{\rotatebox{270}{\includegraphics[width=8cm]{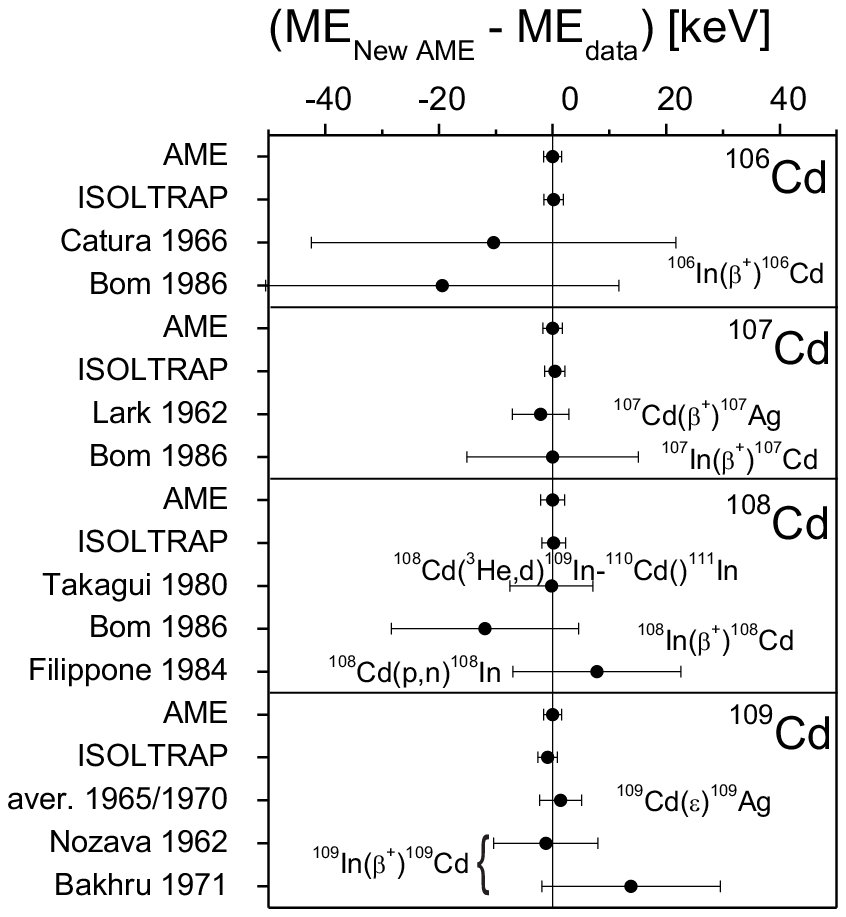}}\label{fig:NEW_AME:Cd107to109}}\qquad
    \caption{\label{fig:NEW_AME:Cd103-109} The difference of the contributing experimental data to the newly evaluated atomic mass-excess is plotted. Note, that the input values might have changed slightly due to feedback from the data of this work. The braces connect the same reaction/experiment.}
  \end{center}
\end{figure}

The comparison of the input data to the new AME value is shown in Fig.\,\ref{fig:NEW_AME:Cd103-109}. Note, that due to feedback from the new data the plotted mass-excess values can shift as compared to Fig.\,\ref{fig:AME2003:Cd99-109}.

\vspace{10mm}
\textbf{$^{109}$Cd}
\newline
The main contribution for the mass-excess value of the AME2003 came from an electron-capture measurement of $^{109}$Cd to $^{109}$Ag with a $Q$-value of 214(3)\,keV as an average of two experiments \cite{Leutz1965,Goedbloed1970} (84.7\%). The other 15.3\% were given by two $\beta^+$-decay $Q$-value measurements $Q$=2015(8)\,keV and 2030(15)\,keV \cite{Nozawa1962,Bakhru1971}. The ISOLTRAP measurement agrees with the earlier values and decreases the experimental uncertainty. After a new evaluation the AME value is now influenced with 82.9\% by the ISOLTRAP data, with 13.7\% by the electron capture $^{109}$Cd(e$^-$)$^{109}$Ag \cite{Leutz1965,Goedbloed1970} and with 3.5\% by $\beta^+$-decay of the $^{109}$In \cite{Nozawa1962,Bakhru1971}.

\vspace{5mm}
\textbf{$^{108}$Cd}
\newline
The mass-excess value of $^{108}$Cd in the AME2003 is calculated including an experimental value from the Minnesota, 16-inch, double-focussing mass spectrometer, namely the difference of $m$(C$_8$H$_{12}-^{108}$Cd$)=189715.6(2.9)$\,$\mu$u \cite{Damerow1963} with 67.9\% influence and the $Q$-value of the  differential reaction of $^{108}$Cd($^3$He,d)$^{109}$In-$^{110}$Cd()$^{111}$In, $Q_0=-806.5(2.6)$\,keV \cite{Takagui1980} (27.1\%). A small contribution comes from the average of a $\beta^+$-decay $Q$-value of $^{108}$In \cite{Bom1986} $Q$=5125(14)\,keV and the $^{108}$Cd(p,n)$^{108}$In reaction \cite{Filippone1984} with a weight of 5.0\%. The ISOLTRAP value compares to the AME2003 within the uncertainties. The result of a new calculation of the AME is determined to 94.0\% by the ISOLTRAP value with a three times smaller uncertainty. The value of the differential reaction \cite{Takagui1980} contributes with 5.7\% and the average of the $\beta$-decay and the (p,n) reaction with 0.3\%.

\vspace{5mm}
\textbf{$^{107}$Cd}
\newline
The mass excess of $^{107}$Cd was determined by the $Q$-value measurements of two $\beta^+$-decays: $Q(^{107}$Cd($\beta^+)^{107}$Ag)\newline =1417(4)\,keV \cite{Lark1962} and $Q(^{107}$In$(\beta^+)^{107}$Cd)=3426(11)\,keV \cite{Bom1986}, which entered with 96.3\% and 3.7\%, respectively, to calculate the AME2003 mass excess. This value agrees with the one from the present work. After reevaluating all data the new AME value is determined to 91.5\% by the ISOLTRAP data. The rest is coming from the $\beta^+$-decay $Q$-values of $^{107}$Cd($\beta^+)^{107}$Ag \cite{Lark1962} and $^{107}$In($\beta^+)^{107}$Cd \cite{Bom1986} with 8.2\% and 0.3\%, respectively.

\vspace{5mm}
\textbf{$^{106}$Cd}
\newline
The mass of $^{106}$Cd was determined by the mass-doublet of C$_8$H$_{10}$-$^{106}$Cd and has been measured to 171789.3(2.7)\,$\mu$u \cite{Damerow1963} contributing to the average value in the AME2003 with 89.0\%. Also the single-neutron pick-up reaction \newline $^{106}$Cd($^3$He,$\alpha)^{105}$Cd ($Q_0=9728(25)$\,keV) \cite{Chapman1975} enters with 4.4\% and the $\beta^+$-decay of $^{106}$In with $Q=6516(30)$\,keV \cite{Catura1966} and $Q=6507(29)$\,keV \cite{Bom1986} combined with the $^{106}$Cd(p,n)$^{106}$In reaction having a reaction $Q$-value of -7312.9(15.0)\,keV \cite{Filippone1984} contribute with 3.5\% to the mass excess value of $^{106}$Cd as tabulated in the AME2003. The measurement at ISOLTRAP agrees with these previous results, but has a four times smaller uncertainty. The new AME result has a 99.7\% influence from the ISOLTRAP data. The $\beta^+$-decay of $^{106}$In \cite{Bom1986,Catura1966} and the (p,n) reaction \cite{Filippone1984} contribute with only 0.3\%. Those have been included due to their significance as links in the mass network.

\vspace{5mm}
\textbf{$^{105}$Cd}
\newline
The two direct mass measurements with the mass excess values of JYFLTRAP and ISOLTRAP agree perfectly within their uncertainties. The previous mass excess value tabled in AME2003 (including experimental data by \cite{Bom1986,Chapman1975,Johnson1953}) is also in agreement within the uncertainties. The new AME value is determined to 92.9\% by ISOLTRAP and to 6.4\% by JYFLTRAP. The $\beta^+$-decay of $^{105}$Cd \cite{Bom1986,Johnson1953} contributes with 0.7\% the mass excess of $^{105}$Cd. In addition, the seven-fold reduction of the uncertainty of $^{105}$Cd mass results also in an improvement of the $ME$ of $^{105}$Ag by a factor of more than two to $\text{-87070.8(4.5)}$\,keV.

\vspace{5mm}
\textbf{$^{104}$Cd}
\newline
For $A=104$ the results of ISOLTRAP and JYFLTRAP agree within the experimental uncertainties, while the SHIPTRAP result deviates from the ISOLTRAP and the JYFL-TRAP value by about 11\,keV (2$\sigma$) and 12\,keV ($1.7\sigma$), respectively. All values agree perfectly with the AME2003, which includes experimental data by \cite{Crawley1982,Dewberry1983,Rotbard1984,Bom1988,Karny1998}, mass excess while reducing the uncertainty. The newly-obtained AME value is to 90.3\% determined by the ISOLTRAP value and by 9.7\% by the JYFLTRAP result.

\vspace{5mm}
\textbf{$^{103}$Cd}
\newline
In this case all three Penning trap measurements agree nicely with each other. Furthermore the three measurements are within the uncertainty of the AME2003 value, which includes experimental data by \cite{Bom1988,Pardo1978}. The newly determined AME value is influenced by 84.7\% and 12.3\% by the ISOLTRAP and the JYFLTRAP value, respectively. There are small contributions coming from the $\beta$-decays $^{103}$Cd($\beta$)$^{103}$Ag (2.4\%) and $^{103}$In($\beta$)$^{103}$Cd (0.6\%).

\vspace{5mm}
\textbf{$^{102}$Cd}
\newline
For this nuclide the mass excess has been also determined at SHIPTRAP and at JYFLTRAP. The two values have a discrepancy of 12\,keV corresponding to 1.4$\sigma$. The ISOLTRAP value agrees well with the measurements at JYFLTRAP but deviates by 1.8$\sigma$ from the values determined with SHIPTRAP. Also in this case all three Penning trap measurements agree with the mass excess listed in the AME2003 which includes experimental data by \cite{Keller1991}. In the new compilation of the mass values for the AME, the ISOLTRAP result contributes with 89.4\% and the JYFLTRAP value with 10.6\%.

\vspace{5mm}
\textbf{$^{101}$Cd}
\newline
The mass excess of $^{101}$Cd has been determined at SHIPTRAP and JYFLTRAP. Both values have a discrepancy of 1.5$\sigma$ relative to each other. The mass excess determined at ISOLTRAP is between the two earlier results, and deviates by 14\,keV (1.3$\sigma$) from the SHIPTRAP results and agrees within the uncertainty with the result from JYFLTRAP. All three values agree with the AME2003 mass-excess determined by \cite{Beck1970,Westgaard1972}. The new AME value is influenced by 92.9\% by the ISOLTRAP result and by 7.1\% by the JYFLTRAP result.

\vspace{5mm}
\textbf{$^{100}$Cd}
\newline
So far, the mass excess of $^{100}$Cd was determined using the SPEG mass spectrometer value of $\text{-}74180(200)$\,keV \cite{Chartier1996} and via the $Q$-value of the $\beta^+$-decay of $^{100}$Cd to $^{100}$Ag of 3890\,keV \cite{Rykaczewski1989}. The experimental result obtained at ISOLTRAP agrees very nicely with the earlier experiments, but the uncertainty is by more than a factor of 50 smaller. The new AME uses the ISOLTRAP data with 100\% of influence for the determination of the $ME$ of $^{100}$Cd, and the connection by $^{100}$Cd$(\beta^+)$$^{100}$In changes the mass excess of $^{100}$In to a mass-excess value of $\text{-}64330(180)$\,keV, indicating that this nucleus is by 35\,keV less bound as compared to the AME2003.

\vspace{5mm}
\textbf{$^{99}$Cd}
\newline
The mass of $^{99}$Cd was determined for the first time by ISOLTRAP. Before, only an AME2003 estimate of the mass excess was available which agrees with the new value determined with ISOLTRAP.

\subsection{Implications for the astrophysical rp process}
\label{sec:astrophysics}

\begin{figure}[ht]
  \begin{center}
      \includegraphics[width=8cm]{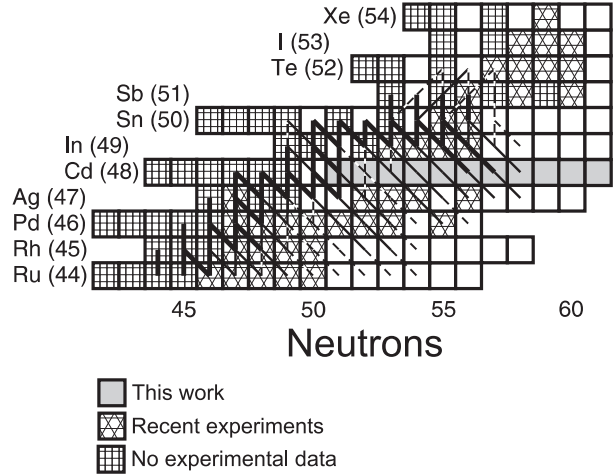}
    \caption{\label{fig:flux-plot} A plot of the time integrated net reaction flows over the entire X-ray burst in the region of the nuclide chart around $^{99}$Cd. The thick lines represent a strong flow (within an order of magnitude of the 3$\alpha$-reaction) and the thin and dashed lines weak flows suppressed by factors of 10 and 100, respectively.  Note that strong proton capture flows either indicate strong net flows, or, due to numerical artefacts, (p,$\gamma$)-($\gamma$,p) equilibrium. The gray shaded nuclides were measured in this work, perpendicularly meshed nuclides represent extrapolated values \protect\cite{Audi2003} and the diagonally meshed boxes indicate nuclides recently measured at other experiments \protect\cite{Fallis2008,Weber:2008pb,Martin2007,Kankainen2008}.}
  \end{center}
\end{figure}

$^{99}$Cd has been suggested as a possible branching point in the path of the astrophysical rp process in some X-ray bursts. Figure~\ref{fig:flux-plot} shows the reaction flows during a type I X-ray burst calculated in a model based on a single-zone approximation and for parameters (accretion rate and initial composition) that are favorable for an extended rp process into the Sn region \cite{Schatz2001,Mazzocchi2007}. Here we updated the reaction network \cite{Cyburt2008} with results from recent Penning trap mass measurements (by e.g. LEBIT \cite{Schury2007}, CPT \cite{Fallis2008}, JYFLTRAP \cite{Weber:2008pb,Kankainen2008} and SHIPTRAP \cite{Weber:2008pb,Martin2007}) and the new masses from this work.

\begin{figure}[ht]
  \begin{center}
      \rotatebox{270}{\includegraphics[width=7cm]{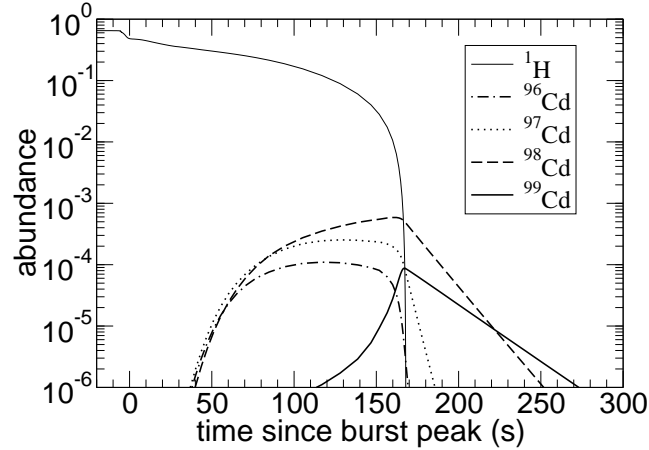}}
    \caption{\label{fig:yt} Abundances of hydrogen and the neutron deficient Cd isotopes as functions of time during an X-ray burst. Zero on the time axis has been chosen to coincide with the burst maximum. The build up of Cd isotopes occurs during the tail of the burst.}
  \end{center}
\end{figure}

Figure~\ref{fig:flux-plot} shows the reaction paths for the entire burst. During the very end of the burst, as hydrogen abundance and temperature are dropping, the reaction path shifts towards $^{99}$Cd (see Fig.~\ref{fig:yt}). The amount of $^{99}$Cd that can be built up by feeding from $^{98}$Cd($\beta^+$)$^{98}$Ag(p,$\gamma$)$^{99}$Cd depends critically on the remaining decrease of $^{99}$Cd by proton captures before hydrogen is completely exhausted and the final abundances freeze out. This depends strongly on the proton separation energy of $^{100}$In, $S_p(^{100}\text{In})$. If this quantity is low, proton captures are inhibited by photo-disintegration of $^{100}$In, and $^{99}$Cd remains abundant as the reaction flow proceeds via its slow $\beta^+$ decay.  If $S_p(^{100}\text{In})$ is large, $^{99}$Cd can be converted very effectively by a dominating reaction flow via $^{99}$Cd(p,$\gamma$)$^{100}$In.

\begin{figure}[ht]
  \begin{center}
      \subfigure[]{\rotatebox{270}{\includegraphics[width=7cm]{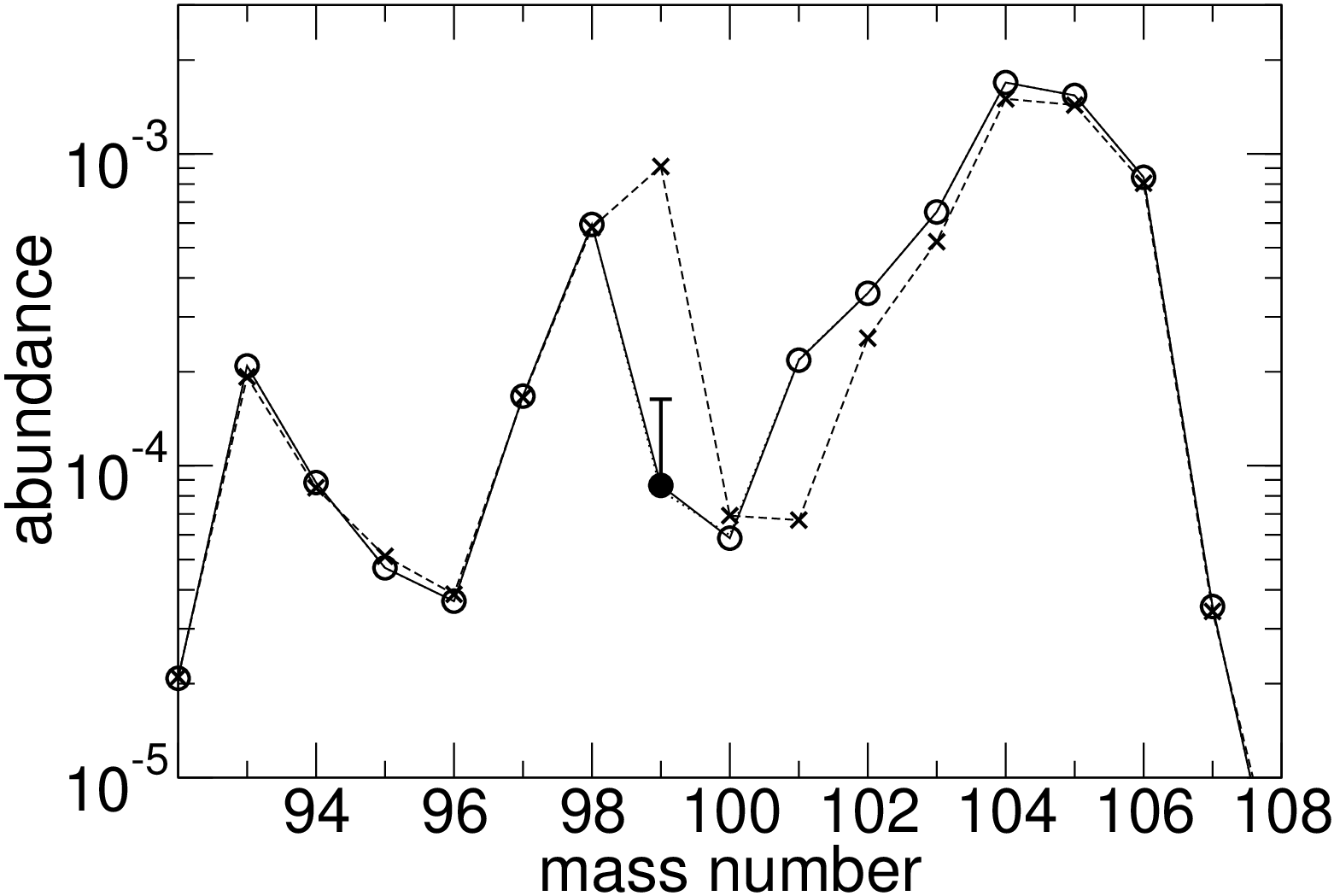}}\label{fig:ashes_ab}}\qquad
      \subfigure[]{\rotatebox{270}{\includegraphics[width=7cm]{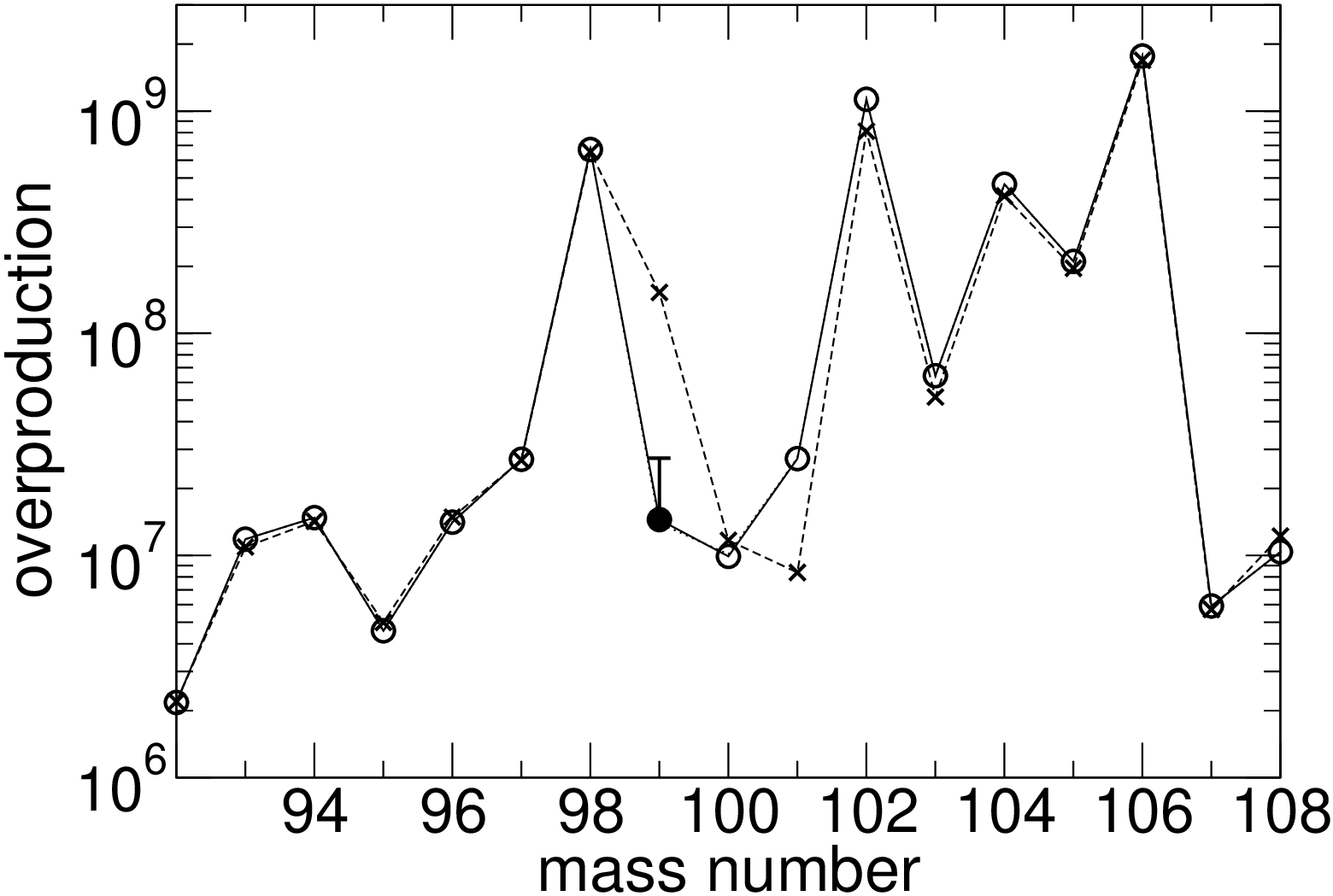}}\label{fig:ashes_ov}}\qquad
    \caption{\label{fig:ashes} \subref{fig:ashes_ab} Final composition of the burst ashes for different values of $S_p(^{100}\text{In})$: with our new value for $^{99}$Cd (circles connected by solid line), the lower 2$\sigma$ limit allowed in AME2003 (crosses connected by dashed line) and the upper 2$\sigma$ limit allowed in AME2003 (dotted line which basically coincides with solid line). The data point with our new $^{99}$Cd mass and our new 2$\sigma$ uncertainty is indicated as a filled circle with error bars. \subref{fig:ashes_ov} Overproduction factors relative to the solar abundance, determined by assuming the entire mass chain has decayed into the first stable isotope. This is a p-nucleus for $A=$92, 94, 96, 98, 102, and 106, while the other mass chains feed isotopes predominantly made by the s process.}
  \end{center}
\end{figure}

The AME2003 value for  $S_p(^{100}\text{In})$ is 1.61(33)\,MeV as obtained adding mass errors quadratically. The large error originated from the extrapolated masses of $^{99}$Cd ($\pm$ 0.21\,MeV) and $^{100}$In ($\pm$ 0.25\,MeV). After our accurate measurement of the $^{99}$Cd mass the uncertainty is almost exclusively due to the $^{100}$In mass. Including the newly evaluated value for the mass of $^{100}$In we obtain now  $S_p(^{100}\text{In})$ of 1.69(18)\,MeV. Figure~\ref{fig:ashes} shows final abundances and overproduction factors relative to solar abundances for model calculations for various values of $S_p(^{100}\text{In})$. Clearly, $S_p(^{100}\text{In})$ is a critical quantity for determining the $A=99$ abundance in the final reaction products (burst ashes). The 2$\sigma$ range of the AME2003 mass uncertainties introduces more than an order of magnitude uncertainty in the $A=99$ abundance.  At the lower 2$\sigma$ limit of  $S_p(^{100}\text{In})$ $A=99$ becomes one of the most abundant mass chains, even exceeding the $A=98$ production by 50\%, while at the upper limit it is one of the least abundant ones. Our new measurements dramatically reduce the possible range of  $A=99$, excluding now an enhanced $A=99$ production at the 2$\sigma$ level. The largest reduction in the uncertainty comes from our precise measurement of the $^{99}$Cd mass. However, the improvement in precision of the mass of $^{100}$In due to the $\beta^+$-decay of $^{100}$In linked to $^{100}$Cd (measured in this work) contributes significantly, leading to an additional reduction of the uncertainty by about a factor of 2.5.

The composition of the burst ashes is important for crust heating models and for judging whether the rp process is a possible production scenario for light p-nuclei. In terms of crustal heating, Gupta  et al. \cite{Gupta2007} have shown that there are significant differences in total heat generation and distribution of heat sources as a function of depth for $A=$98, 99, or 100 ashes. While a change in a single mass chain probably has only a small effect on the thermal structure of the neutron star, our work shows that there are very large uncertainties in the prediction of the final composition of the burst ashes that need to be addressed. Our measurement is a first step in that direction. Uncertainties in other mass chains will also have to be addressed.

In order to judge the suitability of a proposed nucleosynthesis scenario to explain the origin of the elements in the solar system, one key aspect is the pattern of overproduction factors, i.e. the ratio of the produced abundances to the solar abundances (see Fig.~\ref{fig:ashes}). The ratio of the overproduction factor of a given isotope to the highest overproduction factor of the pattern (or to an average of the highest overproduction factors when taking into account variations due to uncertainties) indicates the fraction of solar system material that could originate at most from this nucleosynthesis site.  For a p-process scenario one would require large, comparable overproduction factors for p-nuclei, and significantly reduced overproduction factors for non-p nuclei. As Fig.~\ref{fig:ashes} shows, the rp process in this particular X-ray burst would be a promising scenario to produce the p-nuclei $^{98}$Ru, $^{102}$Pd and $^{106}$Cd. However, co-production of non-p nuclei such as isotopes fed by the $A=99$, 104, and 105 mass chains potentially limits this scenario. The question is whether a possible co-production can be attributed to uncertainties in the nuclear physics, or whether it is a fundamental issue with the proposed scenario. For the $A=99$ case, we have now addressed this question with the present measurement. As Fig.~\ref{fig:ashes} shows, at the 2$\sigma$ level the AME2003 mass uncertainties allowed for co-production of as much as 20\% of $^{99}$Ru (a s-process nucleus) relative to the p-nucleus $^{98}$Ru. With our new mass measurements, $A=99$ co-production is now limited to a rather insignificant few \%.

\section{Conclusion and Outlook}
\label{sec:conclusion}

In the present work, mass determinations of the eleven neutron-deficient nuclides $^{99-109}$Cd are reported. Due to clean production of these nuclides it was possible to reduce the experimental uncertainties down to 2\,keV. In the case of $^{99}$Cd the mass was determined for the first time and for the nuclide $^{100}$Cd the uncertainty was reduced by a factor of more than 50. In addition the influence of the present results on the mass network of the atomic mass evaluation is described as well as the role of the evaluation for solving conflicts of mass data as in the case of $^{101-105}$Cd measured at ISOLTRAP, SHIPTRAP and JYFLTRAP.

The presented mass measurements are an important step towards an understanding of the nuclear physics of the rp process that will enable a more reliable determination of the composition of the produced material at $A=99$. It was shown that the mass of $^{99}$Cd strongly affects the $A=99$ production in a X-ray burst model, and that uncertainties have been significantly reduced from more than an order of magnitude to less than a factor of 2, with the remaining uncertainty coming from the mass of $^{100}$In.

In principle, other uncertainties will also contribute at this level. These include those of masses of lighter Cd isotopes, where similar rp-process branchpoints occur and which might affect feeding into the $^{99}$Cd branchpoint. In addition, nuclear reaction rate uncertainties will also play a role. However, as reaction rates affect branchings in a linear fashion, while mass differences enter exponentially, mass uncertainties will tend to dominate \cite{Schatz2006}. Also, which reaction rates are important depends largely on nuclear masses. For example, for low $S_p(^{100}\text{In})$ a (p,$\gamma$)-($\gamma$,p) equilibrium will be established between $^{99}$Cd and $^{100}$In and the $^{100}$In(p,$\gamma$) reaction rate would affect the $A=99$ production, while for larger $S_p(^{100}\text{In})$ the $^{99}$Cd(p,$\gamma$) reaction rate might be more relevant. Therefore, the mass uncertainties should be addressed first. Once they are under control, further improvements might be possible by constraining proton capture rates.

Our results are relevant for any rp-process scenario with a reaction flow through the $^{99}$Cd
region. Here, we used an X-ray burst model, to investigate in detail the impact of our measurements on
such an rp process. The $\nu$p process in core collapse supernovae might be another
possible scenario for an rp process in the $^{99}$Cd region. It is planned to also explore
whether in that case mass uncertainties have a similar impact on the final composition.

\section*{Acknowledgements}
This work was supported by the German Federal Ministry
for Education and Research (BMBF) (06GF186I, 06MZ215),
the French IN2P3, and the EU FP6 programme (MEIF-CT-2006-042114 and EURONS DS project/515768 RIDS), the Helmholtz Association for National Research Centers (HGF) (VH-NG-037). H.S. is supported by NSF grants PHY0606007 and PHY0216783.  We are grateful to the members of the ISOLDE technical group for their support. Finally we thank V.\,V. Elomaa and the JYFLTRAP group for providing their data prior to publication.

\end{document}